# NETWORK ANALYSIS METHOD: CORRELATION VALUES BETWEEN TWO ARBITRARY POINTS ON A NETWORK

**Akira Saito**

*630-8306 Nara Prefecture Nara City Kidera Cho 649 202*

*saito@tsunagari-llc.jp*

Tsunagari LLC Japan

**Abstract –**This study presents a generalization for a method examining the correlation function of an arbitrary system with interactions in an Ising model to obtain a value of correlation between two arbitrary points on a network. The establishment of a network clarifies the type of calculations necessary for the correlation values between secondary and tertiary nodes. Moreover, it is possible to calculate the correlation values of the target that are interlinked in a complex manner by proposing a network analysis method to express the same as a network with mutual linkages between the target of each field.

## 1. FOREWARD

### From an Ising model to a general network

The analysis method of this network is obtained by generalizing a unique solution for the Ising model in statistical mechanics. The network analysis method referred to here is a solving method that evaluates how to obtain the correlation values of secondary and tertiary nodes prior to a network with a network that includes the same as edges when the correlation between the nodes is known. The method is characterized by the fact that because a network that is generalized and abstracted in scope can be expressed as a network, it can constitute the basis of network theory.

## 2. THEORY

### The Ising model Hamiltonian

An Ising model is constructed of nodes (electrons) with two states, namely, upward spin and downward spin. An interaction energy exists between the two nodes (between two electrons) and is expressed by the following Hamiltonian expression:

$$\mathcal{H}_{i,j} = -J\,\sigma_i\,\sigma_j \qquad \text{Interaction between i and j}$$

,(1)

where $\sigma$ corresponds to +1 when the spin is upward and −1 when the spin is downward.

**Forming a network using Ising model interactions**

The preparation for applications to networks with arbitrary formats is preceded by considering an Ising model of a network with interactions of the previous Ising model as the edges and each electron pointing upwards and downwards as the nodes. The nodes that possess an edge between them are arbitrary. Thus, the system is expressed by the following Hamiltonian:

$$\mathcal{H} = \Big(\sum_{\{i,j\}} -J_{i,j}\,\sigma_i\,\sigma_j\Big) \qquad \text{Interaction between i and j with edges}$$

, (2)

The size of interaction between edges is considered to be different with respect to each edge. Let's obtain the distribution function of the Ising model with the Hamiltonian of this interaction. The distribution function is defined by the following equation according to statistical mechanics:

$$Z = \sum_{\{\sigma\}} \exp(-\mathcal{H} / k_B T)$$

, (3)

where kB denotes the Boltzmann constant, T denotes temperature, and { $\sigma$ } denotes the sum of all spin status of the system. The form expressed by a Hamiltonian of equation (2) is as follows:

$$Z = \sum_{\{\sigma\}} \prod_{\{i,j\}} \exp(J_{i,j}\,\sigma_i\,\sigma_j / k_B T)$$

, (4)

**General representation of an arbitrary format Ising model distribution function**

The following can occur between the nodes (i, j):

(Upward, downward) ＝ (exp(+J/kT))

(Upward, downward) = (exp(−J/kT))

(Downward, upward) = (exp(−J/kT ))

(Downward, upward) = (exp(+J/kT ))

Hence, the aforementioned transformations can be expressed as follows:

$$\sum_{\{\sigma\}}\prod_{\{i,j\}} \exp( J_{i,j}\sigma_i\ \sigma_j / k_B T)$$

$$\Downarrow$$

$$\sum_{\{\sigma\}}\prod_{\{i,j\}} \frac{1}{2}\{\exp( J_{i,j} / k_B T) + \exp( -J_{i,j} / k_B T)\}$$
$$+ \frac{1}{2}\sigma_i\ \sigma_j\ \{\exp( J_{i,j} / k_B T) - \exp( -J_{i,j} / k_B T)\}$$

, (5)

If a new variable Phi and a hatted Phi is introduced, then the following expression is obtained:

$$\phi_{i,j} = \frac{1}{2}\{\exp( J_{i,j} / k_B T) + \exp( -J_{i,j} / k_B T)\}$$
$$\hat{\phi}_{i,j} = \frac{1}{2}\{\exp( J_{i,j} / k_B T) - \exp( -J_{i,j} / k_B T)\}$$

, (6)

$$Z = \sum_{\{\sigma\}}\prod_{\{i,j\}} \{ \phi_{i,j} + \sigma_i\ \sigma_j\ \hat{\phi}_{i,j} \}$$

, (7)

**Route and loop**

A route and a loop are defined to introduce the concept of routes and loops. A route corresponds to the path from a node (electron) to a node (electron) via an edge (interaction).

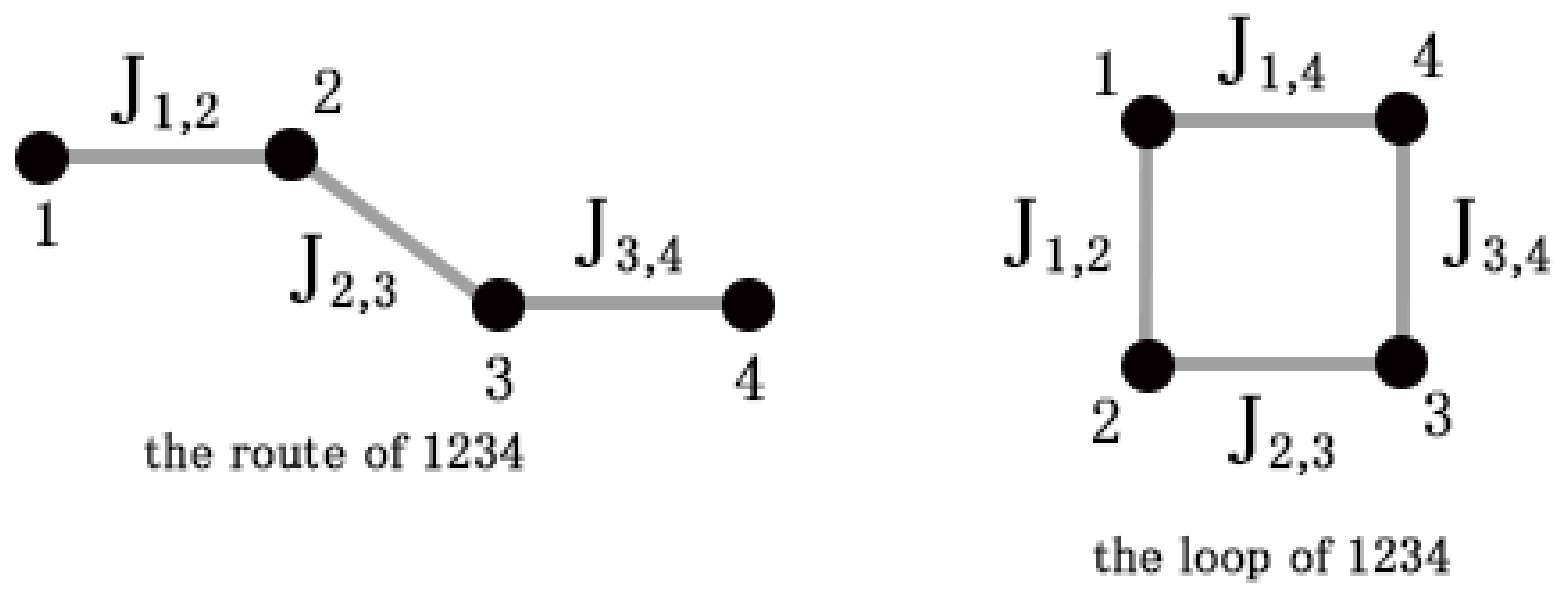

, (8)

Thus, an edge can be passed only once, and it is not permitted to pass twice. A route with the same start and end point corresponds to a loop. Specifically, the minimum distant loop, such as a square loop of a two-dimensional lattice, is defined as a unit loop.

**Representation via distribution function loops**

Equation (7) is expressed as follows:

$$Z = \sum_{\{\sigma\}} \prod_{\{i,j\}} \{ \phi_{i,j} + \sigma_i \sigma_j \hat{\phi}_{i,j} \}$$

, (7)

Within the {} brackets, the hatted φ includes the product of σi and σj, σ corresponds to a value of +1 or −1, and he sum of σ in a value of +1 or −1 is considered across all combinations of σ. Hence, the odd-number terms of σ do not correspond to 1 when the squared term disappears in the summation of the combinations. In addition, Φ and hatted φ represent the interactions between i and j; thus, the remaining even number terms of σ do not include a start and end point. That is, they correspond to loops of hatted φ. Furthermore, a sum is considered for all combinations, and the terms that remain without disappearing correspond to the combinations of all the loops.

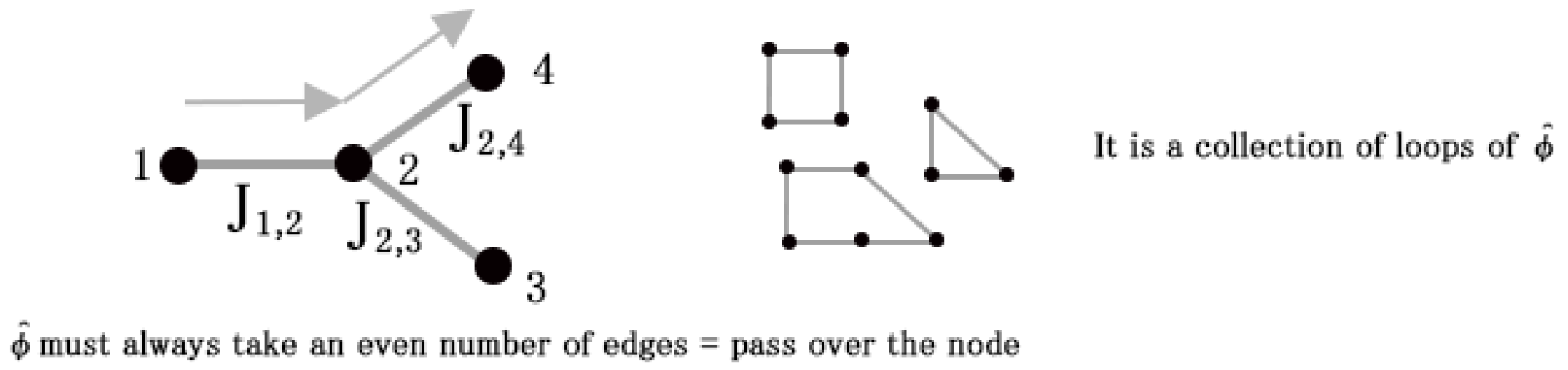

, (9)

$$Z = \{\text{the sum of the collection of loops of } \hat{\phi}\}$$

, (10)

For example, the distribution function of a system is obtained as follows:

the

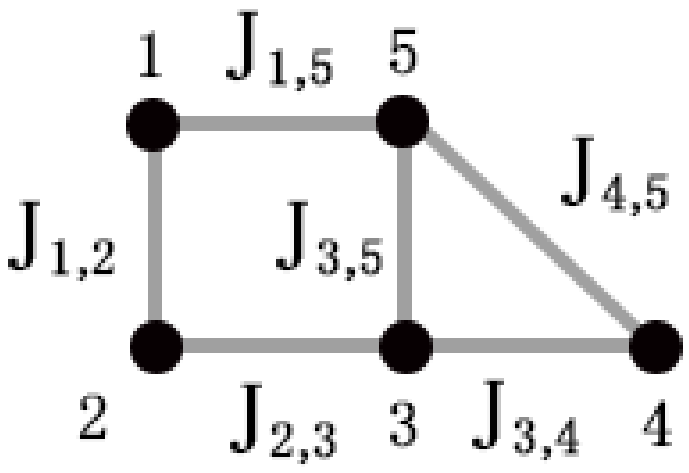

$$\begin{aligned} Z = 2\{ & \phi_{1,2}\phi_{2,3}\phi_{3,4}\phi_{3,5}\phi_{4,5}\phi_{1,5} \quad \text{No loop} \\ & + \hat{\phi}_{1,2}\hat{\phi}_{2,3}\phi_{3,4}\hat{\phi}_{3,5}\phi_{4,5}\hat{\phi}_{1,5} \\ & + \phi_{1,2}\phi_{2,3}\hat{\phi}_{3,4}\hat{\phi}_{3,5}\hat{\phi}_{4,5}\phi_{1,5} \\ & + \hat{\phi}_{1,2}\hat{\phi}_{2,3}\hat{\phi}_{3,4}\phi_{3,5}\hat{\phi}_{4,5}\hat{\phi}_{1,5}\} \end{aligned}$$

, (11)

**Representation by using a combination of correlation function routes and loops**

The correlation function between spins a and b can be expressed based on equation (7) as follows:

$$\langle a,b \rangle = \frac{1}{Z} \sum_{\{\sigma\}} \sigma_a \, \sigma_b \prod_{\{i,j\}} \{ \phi_{i,j} + \sigma_i \, \sigma_j \, \hat{\phi}_{i,j} \}$$

, (12)

A thought process similar to the one utilized while seeking the distribution function is followed. Specifically, σ corresponds to either +1 or −1. A sum is considered for all combinations; therefore, the remaining item corresponds to an even number term of σ. The distribution function is comprehended by understanding all the combinations of the loops. The correlation function is with respect to the σ of node a and node b. Therefore, node a and node b include an odd number of hats φ that connect to them, and it is observed that they correspond to the start and end points. Hence, this is considered as a route from a to b. Sums are considered for all combinations of σ; thus, they evidently correspond to all the combinations of routes from a to b and other nodes.

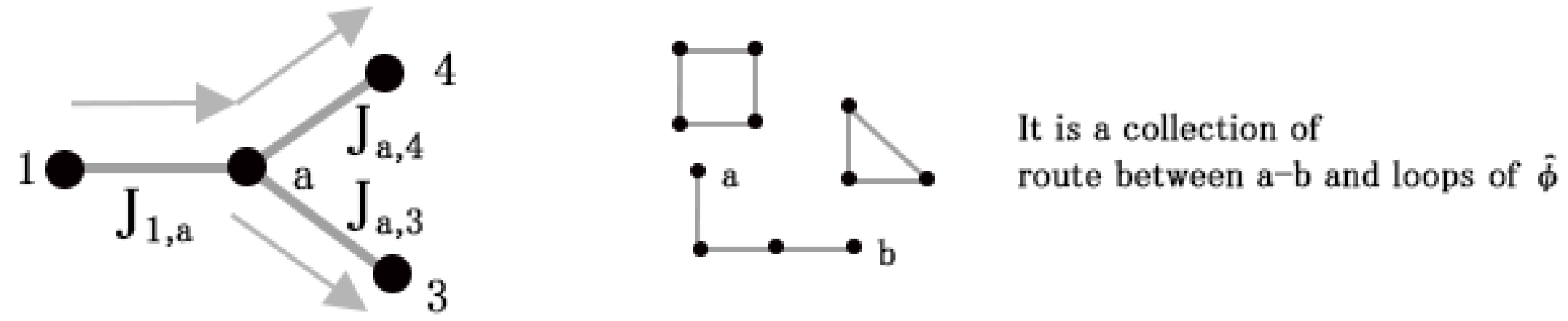

(a) node has an odd number of $\hat{\phi}$ = (a) node is the start and end point of the route

, (13)

$$\langle a,b \rangle = \frac{1}{Z}\{\text{the sum of the collection of a–b route and loops of } \hat{\phi}\}$$

, (14)

For example, the correlation function between a–b of a system similar to the one shown below can be determined as follows:

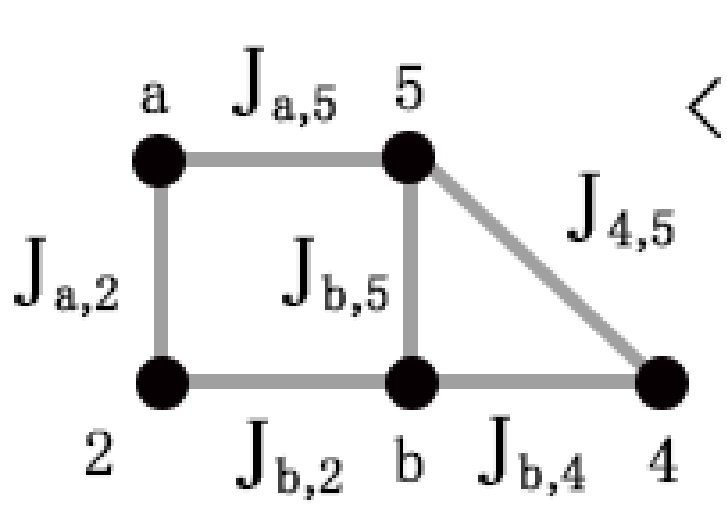

$$\langle a,b \rangle = \frac{1}{Z}\{\ \hat{\phi}_{a,2}\hat{\phi}_{b,2}\phi_{b,4}\phi_{b,5}\phi_{4,5}\phi_{a,5} + \phi_{a,2}\phi_{b,2}\phi_{b,4}\hat{\phi}_{b,5}\phi_{4,5}\hat{\phi}_{a,5} + \phi_{a,2}\phi_{b,2}\hat{\phi}_{b,4}\phi_{b,5}\hat{\phi}_{4,5}\hat{\phi}_{a,5} + \hat{\phi}_{a,2}\hat{\phi}_{b,2}\hat{\phi}_{b,4}\hat{\phi}_{b,5}\hat{\phi}_{4,5}\phi_{a,5}\ \}$$

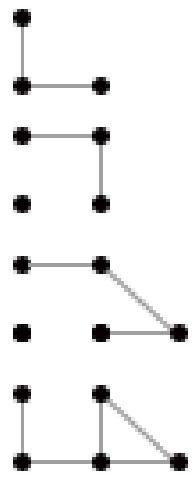

, (15)

**Introduction of the expected value of function A**

Here function A is introduced and is defined as follows:

$$A_{i,j} = \frac{\hat{\phi}_{i,j}}{\phi_{i,j}}$$

, (16)

Equations (7) and (12) are expressed as follows:

$$Z = \left(\prod_{\{i,j\}} \phi_{i,j}\right) \sum_{\{\sigma\}} \prod_{\{i,j\}} \{\ 1 + \sigma_i\ \sigma_j\ A_{i,j}\ \}$$

, (17)

$$\langle a,b \rangle = \frac{1}{Z} \left(\prod_{\{i,j\}} \phi_{i,j}\right) \sum_{\{\sigma\}} \sigma_a \sigma_b \prod_{\{i,j\}} \{ 1 + \sigma_i \sigma_j A_{i,j} \}$$

, (18)

Therefore, the correlation function of specific nodes a and b is as follows:

$$\langle a,b \rangle = \frac{\{\text{the sum of the collection of a–b route and loops of } A_{i,j}\}}{\{\text{the sum of the collection of loops of } A_{i,j}\}}$$

, (19)

For example, the correlation function between a and b corresponds to a system that is expressed as follows:

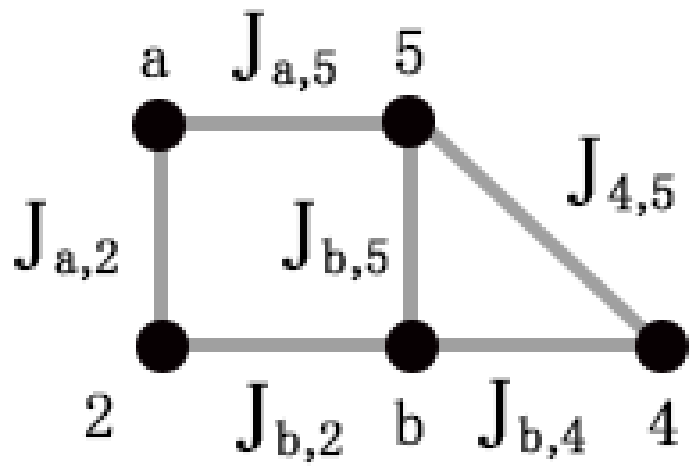

$$\langle a,b \rangle = \frac{\{ A_{a,2} A_{b,2} + A_{a,5} A_{b,5} + A_{b,4} A_{4,5} A_{a,5} + A_{a,2} A_{b,2} A_{b,4} A_{b,5} A_{4,5} \}}{\{ 1 + A_{a,2} A_{a,5} A_{b,2} A_{b,5} + A_{b,4} A_{b,5} A_{4,5} + A_{a,2} A_{b,2} A_{b,4} A_{4,5} A_{a,5} \}}$$

No loop

, (20)

**Expansion to general network analysis method**

Incidentally, A is as defined by equation (16) as follows:

$$A_{i,j} = \frac{e^{+J_{i,j}/k_B T} - e^{-J_{i,j}/k_B T}}{e^{+J_{i,j}/k_B T} + e^{-J_{i,j}/k_B T}}$$

, (21)

where A denotes the correlation function between two nodes based on statistical mechanics. Hence, if the correlation between these two nodes is calculated as per equation (19), then the correlation function between a and b in the system can be obtained. Therefore, irrespective of the form of the system, if the correlation function between two parties with an edge is sought and the sum of products of loops and routes of equation (19) is taken, then the correlation function between arbitrary two points is obtained.

A further generalization is as follows. It is assumed that A denotes a correlation function based on statistics from statistical physics. Equation (21) shows this type of a correlation value. However, the correlation between an arbitrary two nodes on the network incorporates the correlation between two parties in statistical physics and is simply obtained by following the network by using the parameter A. Hence, irrespective of the content of correlation A, if A corresponds to a parameter that represents the correlation between two parties, then the correlation between two arbitrary nodes can be calculated by applying a method similar to equation (19). Hence, it is evident that the correlation between two points on a network is obtained by following equation (19) irrespective of whether the correlation corresponds to statistical dynamics, Bayes statistics, other statistics, or whether it is within A.

Given the above points and a network in which correlation between the edges is indicated by values and functions, it is possible to numerically calculate the manner in which various nodes within a network are interlinked by performing a calculation such as equation (19) that corresponds to an example of equation (20). This corresponds to a method of analyzing the type of networks that comprise the nodes and edges.

**How to obtain network loops**

Finally, the method by which the loops are obtained will be discussed. To perform a calculation, such as equation (19) or the example in equation (20), it is necessary to obtain various loops of the system. However, there are simple ways to obtain various loops of a system. This method involves multiplying all the unit loops with those with no vortex in A and those with a vortex in A. Hence, if the unit loop shares an edge with an adjacent unit loop and both have vortices, then A on the shared section is said to offset each other. It is considered to correspond to a state in which the flow of adjacent vortices stops due to opposing flows that collide on the shared edge. Under the rule that the square of A on the same edge becomes 1, It is possible to calculate all the loops in the system if all the unit loops which is the sum of presence and absence of vortices are multiplied together.

For example, the correlation <1, 3>, and between 1 and 3 in a system is obtained as follows:

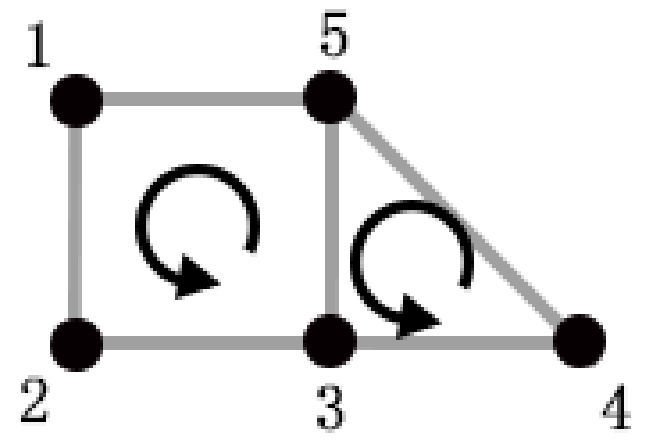

$$\overbrace{(\underset{\text{no vortex}}{1} + \underset{\text{vortex}}{A_{1,2}A_{2,3}A_{3,5}A_{5,1}})}^{\text{unit loop}}\overbrace{(\underset{\text{no vortex}}{1} + \underset{\text{vortex}}{A_{3,4}A_{4,5}A_{5,3}})}^{\text{unit loop}}$$

$$\{\underset{\text{No loop}}{1} + A_{1,2}A_{2,3}A_{3,5}A_{5,1} + A_{3,4}A_{4,5}A_{5,3} + A_{1,2}A_{2,3}A_{3,4}A_{4,5}A_{5,1}\}$$

, (22)

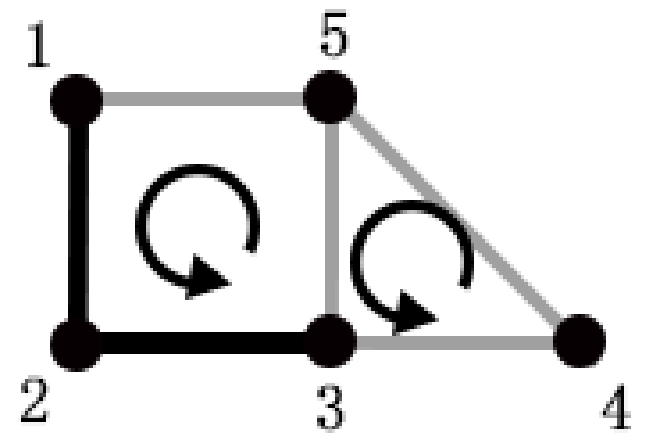

$$A_{1,2}A_{2,3}\overbrace{(\underset{\text{no vortex}}{1} + \underset{\text{vortex}}{A_{1,2}A_{2,3}A_{3,5}A_{5,1}})}^{\text{unit loop}}\overbrace{(\underset{\text{no vortex}}{1} + \underset{\text{vortex}}{A_{3,4}A_{4,5}A_{5,3}})}^{\text{unit loop}}$$

$$\{A_{1,2}A_{2,3} + A_{3,5}A_{5,1} + A_{1,2}A_{2,3}A_{3,4}A_{4,5}A_{5,3} + A_{3,4}A_{4,5}A_{5,3}\}$$

, (23)

$$\langle 1,3\rangle = \frac{\{A_{1,2}A_{2,3} + A_{3,5}A_{5,1} + A_{1,2}A_{2,3}A_{3,4}A_{4,5}A_{5,3} + A_{3,4}A_{4,5}A_{5,3}\}}{\{\underset{\text{No loop}}{1} + A_{1,2}A_{2,3}A_{3,5}A_{5,1} + A_{3,4}A_{4,5}A_{5,3} + A_{1,2}A_{2,3}A_{3,4}A_{4,5}A_{5,1}\}}$$

, (24)

Generalizing the expression yields the following equation:

$$\langle a,b\rangle = \frac{\prod_{\{\text{one a-b route}\}} A_{i,j} \prod_{\{\text{all unit loops}\}} \{1 + \prod_{\{\text{unit loop edge}\}} A_{i,j}\}}{\prod_{\{\text{all unit loops}\}} \{1 + \prod_{\{\text{unit loop edge}\}} A_{i,j}\}}$$

, (25)

However, the following rule applies:

$$A_{i,j}^{2} = 1$$

, (26)

Given the above expression, the rule from equation (26) should evidently be applied to calculate the product of all unit loops (without vortex + with vortex) to analyze a network.

## 3. RESULTS

It is possible to generalize the arbitrary network shaped correlation function with Ising model interaction and to analyze the network made up of edges represented by correlation A based on arbitrary statistics. The correlation between the two nodes of the network is given by the following equation:

$$\langle a,b \rangle = \frac{\{\text{the sum of the collection of a–b route and loops of } A_{i,j}\}}{\{\text{the sum of the collection of loops of } A_{i,j}\}}$$

, (19)

Moreover, it is evident that under the rule that the square of a unit loop and A correspond to one, the following expression is obtained:

$$\langle a,b \rangle = \frac{\prod_{\{\text{one a–b route}\}} A_{i,j} \prod_{\{\text{all unit loops}\}} \{ 1 + \prod_{\{\text{unit loop edge}\}} A_{i,j} \}}{\prod_{\{\text{all unit loops}\}} \{ 1 + \prod_{\{\text{unit loop edge}\}} A_{i,j} \}}$$

, (25)

$$A_{i,j}^{2} = 1$$

, (26)

Specifically, A can correspond to any value given that it represents the correlation between two parties on the edge based on the target statistical method. Hence, all types of links on the network with known correlations between targeted node groups as edges are obtained via equations (19) and (25).

## 4. DISCUSSION

This study aids in the formulation and quantification of the manner in which nodes that are far apart are related simply based on the format of the network by considering statistical methods and individual characteristics. Furthermore, it is possible to analyze the manner in which targets that are linked in a complicated manner from the network composed of relationships between two parties without issues due to the target field or characteristics. Therefore, this could correspond to the method of analyzing network structures with the aforementioned elements in a field of complex systems in which elements that compose the system are interlinked.